\documentclass[preprint,aps,showpacs]{revtex4}
\usepackage[pdftex]{graphicx}
 \begin{document}
  \newcommand{\Qed}{\rule{2.5mm}{3mm}}
 \newcommand{\balpha}{\mbox{\boldmath {$\alpha$}}}
 \newcommand{\mali}[1]{{\scriptscriptstyle#1}}
 \newcommand{\NP}{\mali{NP}}
  \newcommand{\SP}{\mali{SP}}
 \def\Tr{{\rm Tr}}
 \def\(#1)#2{{\stackrel{#2}{(#1)}}}
 \def\[#1]#2{{\stackrel{#2}{[#1]}}}
 \def\A{{\cal A}}
 \def\B{{\cal B}}
 \def\Sb#1{_{\lower 1.5pt \hbox{$\scriptstyle#1$}}}

\title{Can  the matter-antimatter asymmetry be easier to understand within the "spin-charge-family-theory",
predicting twice four families and two times $SU(2)$ vector gauge and scalar fields?  }

\author{ N. S. Manko\v c Bor\v stnik}
\address{ Department of Physics, FMF, University of
Ljubljana, Jadranska 19, 1000 Ljubljana}

\begin{abstract}
This contribution is an attempt to try to understand the matter-antimatter asymmetry in the universe
within the  {\it spin-charge-family-theory}~\cite{norma,pikanorma}  if assuming that  
transitions in non equilibrium processes among instanton vacua and complex phases in  
mixing matrices are the sources of the matter-antimatter asymmetry, as 
studied in the literature~\cite{gross,rubakovshaposhnikov,dinekusenko,tapeiling} 
for several proposed theories. 
The {\it spin-charge-family-theory} is, namely, very   
promising in showing the right way beyond  the {\it standard model}.  
It predicts  families and their mass matrices, explaining  the origin of the charges
and of the gauge fields.  It predicts that there are, after the  universe passes through two 
$SU(2)\times U(1)$ phase transitions,
in which the symmetry  breaks from $SO(1,3) \times SU(2) \times SU(2) \times U(1) \times SU(3)$ first 
to $SO(1,3) \times SU(2)  \times U(1) \times SU(3)$ and then to 
   $SO(1,3) \times U(1) \times SU(3)$,  twice decoupled four families. The upper four families gain masses 
   in the first phase transition, while the second four families gain masses at the electroweak break. 
To these two breaks of symmetries the scalar non Abelian fields, the (superposition of the) 
gauge fields of the operators generating  families, contribute.
The lightest of the upper four families 
is stable (in comparison with the life of the universe) and is therefore a candidate for 
constituting the dark matter. The heaviest of the lower four families should be seen at the LHC or 
at somewhat higher energies.  
\end{abstract}


\maketitle

\date{\today}

\section{Introduction}
\label{introduction}

The {\it theory unifying spin and charges and predicting families} ({\it spin-charge-family-theory}) 
assumes that spinors carry in $d \ge 4$ ($d= 1 + 13$ is studied) only two kinds of the spin. 
The Dirac kind  $\gamma^{a}$  manifests after  several appropriate breaks of the 
starting symmetry as the spin and all the charges. 
The second kind called  $\{ \gamma^a$ ($\{ \gamma^a, \tilde{\gamma}^b\}_{+}=0$) generates families. 
 Accordingly there 
 are in $d \ge 4$, besides the vielbeins, also the two kinds of the spin connection fields,  which 
 are the gauge fields of the corresponding operators  $S^{ab}$ and $\tilde{S}^{ab}$.  Those 
 connected with $S^{ab}$ manifest in $d=(1+3)$ as the vector gauge fields, while those 
 connected with $\tilde{S}^{ab}$ manifest as the scalar fields and determine on the tree level the 
 mass matrices.

Let me make a short review  of the so far made predictions of the 
{\it spin-charge-family-theory}:

\begin{itemize}

\item   The {\it spin-charge-family-theory} has the explanation for 
        the appearance of the internal degrees of freedom -- the  spin   
        and the charges while unifying them under the assumption that the  
        universe went through several phase transitions which cause the appropriate breaks 
        of the starting symmetry. Then the fact that the right handed (with respect to SO(1,3)) 
        fermions are weak chargeless, while the left handed ones carry the weak charge emerges, 
        as well as that there exist leptons (singlets with respect to the 
        colour charge) and quarks (triplets with respect to the colour charge)~\cite{norma,pikanorma}. 
        
\item   The theory  explains the appearance of massless families at 
        the low energy regime under the 
        assumption that there are breaks which leave the massless fermions of only one 
        handedness~\cite{hnd}. Assuming that breaks of symmetries affect 
        the whole internal space --- the space defined by both kinds of the Clifford algebra objects ---         
        it predicts in the energy regime close below $10^{16}$ GeV 
        eight massless families. The manifested symmetry is (assumed to be) at this stage 
        $SO(1,3) \times SO(4) \times U(1) \times SU(3)$. 
        The next  break of  the symmetry of the universe to $SO(1,3) \times SU(2) \times 
        U(1) \times SU(3)$ leaves four families massless~\cite{pikanorma}, while  the vacuum 
        expectation values of superposition of the  starting fields which manifest in $(1+3)$ 
        as  scalar fields, 
        make the upper four families  and the corresponding gauge fields massive. 
        After the electroweak break also the lower four 
        families become massive due to the vacuum expectation values of superposition of 
        the starting fields, together with the weak bosons. 
        
\item   The theory predicts the fourth family, which will be observed 
        at the LHC or at somewhat higher energies~\cite{gmdn}, 
        and the fifth stable family (with no mixing matrix elements couplings 
        to the lower four families in comparison with the age of the universe), the baryons and 
        neutrinos of which are the candidates to form the dark matter.

\item   The masses of this fifth family members are according to the so far made rough 
        estimations~\cite{pikanorma,gmdn} larger than a few TeV and smaller than $10^{10}$  TeV. 
        The members of the family have approximately the same mass, 
        at least on the tree level~\cite{normaproc2010talk}. 
        
\item   The studies~\cite{gn}  of the history of the stable fifth family members in the evolution 
	of the universe and  of their interactions with the ordinary matter in the DAMA's and the 
	CDMS's experiments done so far lead to the prediction that the masses of the fifth family 
	members, if they constitute the dark matter, 
        are  a few hundred TeV, independent of the fifth family fermion-antifermion asymmetry. 
        The Xe experiment looks like to be in disagreement, but careful analyses show that one 
        should wait for further data~\cite{discussGN} to make the final conclusion. 
        
        The lightest fifth family baryon is, in the case that all the quarks have approximately 
        (within a hundred GeV) the same mass~\cite{gn},  the fifth 
        family neutron, due to the attractive electromagnetic interaction. The difference in the 
        weak interaction can be  for large enough masses neglected.
         
\item	The fermion asymmetry  in the {\it approach} has not yet really been studied.

\item   The studies~\cite{gn}  of the evolution of this stable fifth family members rely  
        on my rough estimations~\cite{gn} of the behaviour of the  coloured  fifth family 
        objects (single quarks and antiquarks  or coloured pairs of quarks or of antiquarks)  
        during the colour phase transition. 
        These estimations namely suggest that the coloured objects either annihilate with the 
        anti-objects or they form  colourless neutrons and antineutrons and correspondingly decouple 
        from the plasma soon after the colour phase transition starts, due to the very strong 
        binding energy of the fifth family baryons (with respect to the first family baryons) 
        long enough before the first family quarks start to form the baryons. 
        These estimations should be followed by more accurate studies. 
        
\item   The so far done studies
        suggest strongly that the number density of the fifth 
        family neutrinos (of approximately the same mass as the fifth family quarks and leptons), 
        which also contribute to the dark matter, is pretty much reduced due to the 
        neutrino-antineutrino annihilation closed below the electroweak break.  
        The weak annihilation cross section is expected to play much stronger role 
        for neutrinos than for strongly bound fifth family quarks in the fifth family neutron 
        (due to the huge binding energy of the fifth family quarks), what also remains to be proved. 

\item   The estimations~\cite{gmdn} of the properties of the lower four families on the tree level 
        call for the calculations beyond the tree level, which should hopefully demonstrate, that 
        the loop corrections (in all orders) bring the main differences in the properties of 
        the family members. These calculations are in progress~\cite{AN}.

        \end{itemize}
 
 Although we can say that the  {\it spin-charge-family-theory} looks  very promising as 
 the right way to  explain where do the assumptions of the {\it standard model} originate, there are 
 obviously many not yet studied, or at least far from being
 carefully enough studied open problems. 
 Many a problem is common to 
 all the theories, like the first family baryon asymmetry, which I am going to discuss within the 
 {\it spin-charge-family-theory}  in this contribution.
  Some of the problems are common  to all the theories  assuming more than so far observed $(1+3)$ 
  dimensions, in particular the {\it spin-charge-family-theory} shares some problems with 
  all the Kaluza-Klein-like theories. We are trying  to solve them first on toy models~\cite{hnd}.  
 
 The main new step in the {\it spin-charge-family-theory} --- the explanation of the appearance of 
 families by assuming that both existing kinds of the Clifford algebra objects should be used
 to treat correctly 
 the fermion degrees of freedom --- limits very much the 
 properties of families and their members. The simple starting action in $d= (1+13)$, which in $d=(1+3)$ 
 demonstrates the mass matrices, namely  fixes to high extent the fermion properties after 
 the breaks of symmetries. Therefore this proposal might soon be studied accurately enough to show whether 
 it is the right theory or not.

 This contribution is an attempt to try to understand  what can the {\it spin-charge-family-theory}  
 say about the  fermion-antifermion asymmetry when taking into account the 
 proposals of the references~\cite{rubakovshaposhnikov,dinekusenko,tapeiling} (and 
 of the works cited therein). These works study the soliton solutions of non Abelian 
 gauge fields  with many different vacua 
  and evaluate fermion number nonconservation due to possible transitions among different  
  vacua in non equilibrium processes 
 during the phase  transitions through which the universe passed. In such processes  
 fermion (and also antifermion) currents are  not conserved since $CP$ is not nonconserved. 
 To the $CP$ nonconservation  also the   complex matrix elements determining the transitions 
 among families contribute and  consequently influence the first family fermion-antifermion   
  asymmetry. 
 
 Since the {\it spin-charge-family-theory} predicts below the unification scale of 
 all the charges two kinds of  phase transitions 
 (first from $SO(1,3) \times SU(2) \times SU(2) \times U(1) \times SU(3)$ to 
 $SO(1,3) \times SU(2) \times U(1) \times SU(3)$, in which the upper four families gain masses and so do 
 the corresponding vector gauge fields, and then from 
 $SO(1,3) \times SU(2) \times U(1) \times SU(3)$  to 
 $SO(1,3) \times U(1) \times SU(3)$, in which the lower four families and the 
 corresponding gauge fields gain masses), in which besides the vector gauge fields
 also  the scalar gauge fields (the gauge fields of $\tilde{S}^{ab}$ and also of $S^{ab}$ 
 with the scalar index with respect to $(1+3)$) contribute,  
 the fermion-antifermion asymmetry might very probably have for the stable fifth family  an opposite sign  
 than for the first family.

 It might therefore  be that the existence of 
  two kinds of four families, together with two kinds of the vector gauge fields and two kinds of the 
  scalar fields help to 
  easier understanding the first family fermion-antifermion asymmetry.

 Although I am studying the fermion asymmetry, together with the discrete symmetries, in   
 the {\it spin-charge-family-theory} for quite some time (not really intensively), this contribution 
 is  stimulated by the question of M.Y. Khlopov~\cite{MYN}, since he is 
 proposing the scenario, in which my stable fifth family members should  manifest an 
 opposite fermion asymmetry than the 
 first family members, that is antifermion-fermion asymmetry. While in the case that the fifth family 
 members have masses around 100 TeV or higher and the neutron is the lightest baryon and 
 neutrino the lightest  lepton~\cite{gn}  the fifth family baryon asymmetry plays no role  
 (since in this case the fifth family neutrons and neutrinos as well as their antiparticles 
 interact weakly enough  among  themselves and with the ordinary matter that the 
 assumption that they constitute the dark matter is in agreement  with the observations). Maxim~\cite{m} 
 claims that the fifth family members with the quark masses not higher than  10 TeV are also the candidates 
 for the dark matter, provided that $\bar{u}_5 \bar{u}_5 \bar{u}_5$ is the lightest antibaryon and   that 
 there is an excess of antibaryons over the baryons in the fifth family case.

 \section{A short overview of the theory unifying spin and charges and explaining families}
 \label{approach}

  In this section I briefly repeat the main ideas of the {\it spin-charge-family-theory}. 
   I kindly ask the reader to learn more about this theory in the references~\cite{norma,pikanorma}  
   as well as in my talk presented in this proceedings and in the references therein. 
 
I am proposing a simple action in $d=(1+13)$-dimensional space. Spinors carry two kinds of the spin 
(no charges).\\ 
i. The Dirac spin, described by $\gamma^a$'s, defines the spinor representation in $d=(1+ 13)$.  
After the break of the starting symmetry  $SO(1,13)$ (through $SO(1,7) \times 
SO(6)$) to the symmetry of the {\it standard model}  in $d=(1+3)$ ($SO(1,3)\times U(1)\times SU(2)\times SU(3)$)  
it defines  the hyper charge ($U(1)$),  the weak charge ($SU(2)$, with the 
left handed representation of $SO(1,3)$ manifesting naturally the weak charge and the right 
handed ones appearing as the weak singlets) and  the colour charge ($SU(3)$). \\
ii. The second kind of the spin~\cite{norma},  
described by $\tilde{\gamma}^a$'s ($\{\tilde{\gamma}^a, \tilde{\gamma}^b\}_{+}= 2 \, \eta^{ab}$) and  
anticommuting with the Dirac $\gamma^a$ ($\{\gamma^a, \tilde{\gamma}^b\}_{+}=0$),  
defines the families of spinors.\\ 
Accordingly spinors interact with the two kinds of the spin connection fields and the vielbeins. 

We have
\begin{eqnarray}
&& \{ \gamma^a, \gamma^b\}_{+} = 2\eta^{ab} =  
\{ \tilde{\gamma}^a, \tilde{\gamma}^b\}_{+},\quad
\{ \gamma^a, \tilde{\gamma}^b\}_{+} = 0,\nonumber\\
&&S^{ab}: = (i/4) (\gamma^a \gamma^b - \gamma^b \gamma^a), \quad
\tilde{S}^{ab}: = (i/4) (\tilde{\gamma}^a \tilde{\gamma}^b 
- \tilde{\gamma}^b \tilde{\gamma}^a),\quad  \{S^{ab}, \tilde{S}^{cd}\}_{-}=0.
\label{snmb:tildegclifford}
\end{eqnarray}
The action
\begin{eqnarray}
S            \,  &=& \int \; d^dx \; E\;{\mathcal L}_{f} +  
\nonumber\\  
               & & \int \; d^dx \; E\; (\alpha \,R + \tilde{\alpha} \, \tilde{R})\,,
               \end{eqnarray}
\begin{eqnarray}
{\mathcal L}_f &=& \frac{1}{2}\, (E\bar{\psi} \, \gamma^a p_{0a} \psi) + h.c.\,, 
\nonumber\\
p_{0a }        &=& f^{\alpha}{}_a p_{0\alpha} + \frac{1}{2E}\, \{ p_{\alpha}, E f^{\alpha}{}_a\}_-, 
\nonumber\\  
   p_{0\alpha} &=&  p_{\alpha}  - 
                    \frac{1}{2}  S^{ab} \omega_{ab \alpha} - 
                    \frac{1}{2}  \tilde{S}^{ab}   \tilde{\omega}_{ab \alpha}\,,                   
\nonumber\\ 
R              &=&  \frac{1}{2} \, \{ f^{\alpha [ a} f^{\beta b ]} \;(\omega_{a b \alpha, \beta} 
- \omega_{c a \alpha}\,\omega^{c}{}_{b \beta}) \} + h.c. \;, 
\nonumber\\
\tilde{R}      &=& \frac{1}{2}\,   f^{\alpha [ a} f^{\beta b ]} \;(\tilde{\omega}_{a b \alpha,\beta} - 
\tilde{\omega}_{c a \alpha} \tilde{\omega}^{c}{}_{b \beta}) + h.c.\;, 
\label{wholeaction}
\end{eqnarray}
manifests ($f^{\alpha [a} f^{\beta b]}= f^{\alpha a} f^{\beta b} - f^{\alpha b} f^{\beta a}$) 
after the break of symmetries all the known 
gauge fields and the scalar fields, and the mass matrices. 
To see the manifestation of the covariant momentum and the mass matrices we rewrite 
formally the action for a Weyl spinor in $d=(1+13)$  as follows  
\begin{eqnarray}
{\mathcal L}_f &=&  \bar{\psi}\gamma^{m} (p_{m}- \sum_{A,i}\; g^{A}\tau^{Ai} A^{Ai}_{m}) \psi 
+ \nonumber\\
               & &  \{ \sum_{s=7,8}\;  \bar{\psi} \gamma^{s} p_{0s} \; \psi \}  + \nonumber\\
               & & {\rm the \;rest}, 
\label{faction}
\end{eqnarray}
where $m=0,1,2,3$ with
\begin{eqnarray}
\tau^{Ai} = \sum_{a,b} \;c^{Ai}{ }_{ab} \; S^{ab},
\nonumber\\ 
\{\tau^{Ai}, \tau^{Bj}\}_- = i \delta^{AB} f^{Aijk} \tau^{Ak}.
\label{tau}
\end{eqnarray}
All the charges and the spin of one family are determined by $S^{ab}$, 
with $S^{ab}$ as the only internal degree of freedom of one family  
(besides the family quantum number, determined by $\tilde{S}^{ab}$),  
manifesting after the breaks at the low energy regime as the generators of the observed 
groups~(Eq.~(\ref{tau})) $U(1), SU(2)$ and $SU(3)$, for $A=1,2,3$, respectively.

The breaks of the starting symmetry from $SO(1,13)$ to the symmetry $SO(1,7) \times SU(3) \times U(1)$ 
and further to $SO(1,3) \times SU(2) \times SU(2) \times U(1) \times SU(3) $ 
are assumed to leave all the low lying families of spinors massless. 
There are eight such massless families ($2^{8/2-1}$) 
before further breaks. 

Accordingly the first row of the action in Eq.~(\ref{faction}) manifests the effective {\it standard model} 
fermions part of the action before the weak break, while the second part manifests, 
after the appropriate breaks of 
symmetries (when $\omega_{ab \sigma}$  and $\tilde{\omega}_{ab \sigma}$, $\sigma \in (5,6,7,8),$ 
fields gain the nonzero vacuum expectation values 
on the tree level)  the mass matrices.

 The generators $\tilde{S}^{ab}$ take care of the families, transforming each member of one family 
 into the corresponding member of another family, due to the fact that 
 $\{S^{ab}, \tilde{S}^{cd}\}_{-}=0$ (Eq.(\ref{snmb:tildegclifford})). 
 
 Using the technique~\cite{snmb:hn02hn03} and analysing the vectors as the eigenvectors of the 
 {\it standard model} groups we present vectors in the space of charges and spins in terms of 
 projectors and nilpotents as can be learned in Appendix, in the references~\cite{norma,pikanorma} and 
 also in my talk in the Proceedings of Bled workshop 2010. 
 
 I present  in Table~\ref{Table I.} the eightplet (the representation of $SO(1,7)$ of quarks of a 
 particular colour charge ($\tau^{33}=1/2$, 
 $\tau^{38}=1/(2\sqrt{3})$), and $U(1)$ charge ($\tau^{4}=1/6$) and on Table~\ref{Table Il.} the 
 eightplet of the corresponding (colour chargeless) leptons. 
 %
 %
 \begin{table}
 \begin{center}
 \begin{tabular}{|r|c||c||c|c||c|c|c||r|r|}
 \hline
 i&$$&$|^a\psi_i>$&$\Gamma^{(1,3)}$&$ S^{12}$&$\Gamma^{(4)}$&
 $\tau^{13}$&$\tau^{23}$&$Y$&$Q$\\
 \hline\hline
 && ${\rm Octet},\;\Gamma^{(1,7)} =1,\;\Gamma^{(6)} = -1,$&&&&&&& \\
 && ${\rm of \; quarks}$&&&&&&&\\
 \hline\hline
 1&$ u_{R}^{c1}$&$ \stackrel{03}{(+i)}\,\stackrel{12}{(+)}|
 \stackrel{56}{(+)}\,\stackrel{78}{(+)}
 ||\stackrel{9 \;10}{(+)}\;\;\stackrel{11\;12}{[-]}\;\;\stackrel{13\;14}{[-]} $
 &1&$\frac{1}{2}$&1&0&$\frac{1}{2}$&$\frac{2}{3}$&$\frac{2}{3}$\\
 \hline 
 2&$u_{R}^{c1}$&$\stackrel{03}{[-i]}\,\stackrel{12}{[-]}|\stackrel{56}{(+)}\,\stackrel{78}{(+)}
 ||\stackrel{9 \;10}{(+)}\;\;\stackrel{11\;12}{[-]}\;\;\stackrel{13\;14}{[-]}$
 &1&$-\frac{1}{2}$&1&0&$\frac{1}{2}$&$\frac{2}{3}$&$\frac{2}{3}$\\
 \hline
 3&$d_{R}^{c1}$&$\stackrel{03}{(+i)}\,\stackrel{12}{(+)}|\stackrel{56}{[-]}\,\stackrel{78}{[-]}
 ||\stackrel{9 \;10}{(+)}\;\;\stackrel{11\;12}{[-]}\;\;\stackrel{13\;14}{[-]}$
 &1&$\frac{1}{2}$&1&0&$-\frac{1}{2}$&$-\frac{1}{3}$&$-\frac{1}{3}$\\
 \hline 
 4&$ d_{R}^{c1} $&$\stackrel{03}{[-i]}\,\stackrel{12}{[-]}|
 \stackrel{56}{[-]}\,\stackrel{78}{[-]}
 ||\stackrel{9 \;10}{(+)}\;\;\stackrel{11\;12}{[-]}\;\;\stackrel{13\;14}{[-]} $
 &1&$-\frac{1}{2}$&1&0&$-\frac{1}{2}$&$-\frac{1}{3}$&$-\frac{1}{3}$\\
 \hline
 5&$d_{L}^{c1}$&$\stackrel{03}{[-i]}\,\stackrel{12}{(+)}|\stackrel{56}{[-]}\,\stackrel{78}{(+)}
 ||\stackrel{9 \;10}{(+)}\;\;\stackrel{11\;12}{[-]}\;\;\stackrel{13\;14}{[-]}$
 &-1&$\frac{1}{2}$&-1&$-\frac{1}{2}$&0&$\frac{1}{6}$&$-\frac{1}{3}$\\
 \hline
 6&$d_{L}^{c1} $&$\stackrel{03}{(+i)}\,\stackrel{12}{[-]}|
 \stackrel{56}{[-]}\,\stackrel{78}{(+)}
 ||\stackrel{9 \;10}{(+)}\;\;\stackrel{11\;12}{[-]}\;\;\stackrel{13\;14}{[-]} $
 &-1&$-\frac{1}{2}$&-1&$-\frac{1}{2}$&0&$\frac{1}{6}$&$-\frac{1}{3}$\\
 \hline
 7&$ u_{L}^{c1}$&$\stackrel{03}{[-i]}\,\stackrel{12}{(+)}|
 \stackrel{56}{(+)}\,\stackrel{78}{[-]}
 ||\stackrel{9 \;10}{(+)}\;\;\stackrel{11\;12}{[-]}\;\;\stackrel{13\;14}{[-]}$
 &-1&$\frac{1}{2}$&-1&$\frac{1}{2}$&0&$\frac{1}{6}$&$\frac{2}{3}$\\
 \hline
 8&$u_{L}^{c1}$&$\stackrel{03}{(+i)}\,\stackrel{12}{[-]}|\stackrel{56}{(+)}\,\stackrel{78}{[-]}
 ||\stackrel{9 \;10}{(+)}\;\;\stackrel{11\;12}{[-]}\;\;\stackrel{13\;14}{[-]}$
 &-1&$-\frac{1}{2}$&-1&$\frac{1}{2}$&0&$\frac{1}{6}$&$\frac{2}{3}$\\
 \hline\hline
 \end{tabular}
 \end{center}
 \caption{\label{Table I.} The 8-plet of quarks - the members of $SO(1,7)$ subgroup of the 
 group $SO(1,13)$, 
 belonging to one Weyl left 
 handed ($\Gamma^{(1,13)} = -1 = \Gamma^{(1,7)} \times \Gamma^{(6)}$) spinor representation of 
 $SO(1,13)$. 
 It contains the left handed weak charged quarks and the right handed weak chargeless quarks 
 of a particular 
 colour $(1/2,1/(2\sqrt{3}))$. Here  $\Gamma^{(1,3)}$ defines the handedness in $(1+3)$ space, 
 $ S^{12}$ defines the ordinary spin (which can also be read directly from the basic vector, both
 vectors  with both spins, $\pm \frac{1}{2}$, are presented), 
 $\tau^{13}$ defines the third component of the weak charge, $\tau^{23}$ the third component 
 of the $SU(2)_{II}$ charge, 
 $\tau^{4}$ (the $U(1)$ charge) defines together with the
 $\tau^{23}$  the hyper charge ($Y= \tau^4 + \tau^{23}$), $Q= Y + \tau^{13}$ is the 
 electromagnetic charge. 
 The reader can find the whole Weyl representation in the ref.~\cite{Portoroz03}.}
 \end{table}
 %
 %
 
 %
 %
 \begin{table}
 \begin{center}
 \begin{tabular}{|r|c||c||c|c||c|c|c||r|r|}
 \hline
 i&$$&$|^a\psi_i>$&$\Gamma^{(1,3)}$&$ S^{12}$&$\Gamma^{(4)}$&
 $\tau^{13}$&$\tau^{23}$&$Y$&$Q$\\
 \hline\hline
 && ${\rm Octet},\;\Gamma^{(1,7)} =1,\;\Gamma^{(6)} = -1,$&&&&&&& \\
 && ${\rm of \; quarks}$&&&&&&&\\
 \hline\hline
 1&$ \nu_{R}$&$ \stackrel{03}{(+i)}\,\stackrel{12}{(+)}|
 \stackrel{56}{(+)}\,\stackrel{78}{(+)}
 ||\stackrel{9 \;10}{(+)}\;\;\stackrel{11\;12}{(+)}\;\;\stackrel{13\;14}{(+)} $
 &1&$\frac{1}{2}$&1&0&$\frac{1}{2}$&$0$&$0$\\
 \hline 
 2&$\nu_{R}$&$\stackrel{03}{[-i]}\,\stackrel{12}{[-]}|\stackrel{56}{(+)}\,\stackrel{78}{(+)}
 ||\stackrel{9 \;10}{(+)}\;\;\stackrel{11\;12}{[-]}\;\;\stackrel{13\;14}{[-]}$
 &1&$-\frac{1}{2}$&1&0&$\frac{1}{2}$&$0$&$0$\\
 \hline
 3&$e_{R}$&$\stackrel{03}{(+i)}\,\stackrel{12}{(+)}|\stackrel{56}{[-]}\,\stackrel{78}{[-]}
 ||\stackrel{9 \;10}{(+)}\;\;\stackrel{11\;12}{[-]}\;\;\stackrel{13\;14}{[-]}$
 &1&$\frac{1}{2}$&1&0&$-\frac{1}{2}$&$-1$&$-1$\\
 \hline 
 4&$ e_{R} $&$\stackrel{03}{[-i]}\,\stackrel{12}{[-]}|
 \stackrel{56}{[-]}\,\stackrel{78}{[-]}
 ||\stackrel{9 \;10}{(+)}\;\;\stackrel{11\;12}{[-]}\;\;\stackrel{13\;14}{[-]} $
 &1&$-\frac{1}{2}$&1&0&$-\frac{1}{2}$&$-1$&$-1$\\
 \hline
 5&$e_{L}$&$\stackrel{03}{[-i]}\,\stackrel{12}{(+)}|\stackrel{56}{[-]}\,\stackrel{78}{(+)}
 ||\stackrel{9 \;10}{(+)}\;\;\stackrel{11\;12}{[-]}\;\;\stackrel{13\;14}{[-]}$
 &-1&$\frac{1}{2}$&-1&$-\frac{1}{2}$&0&$-\frac{1}{2}$&$-1$\\
 \hline
 6&$e_{L} $&$\stackrel{03}{(+i)}\,\stackrel{12}{[-]}|
 \stackrel{56}{[-]}\,\stackrel{78}{(+)}
 ||\stackrel{9 \;10}{(+)}\;\;\stackrel{11\;12}{[-]}\;\;\stackrel{13\;14}{[-]} $
 &-1&$-\frac{1}{2}$&-1&$-\frac{1}{2}$&0&$-\frac{1}{2}$&$-1$\\
 \hline
 7&$ \nu_{L}$&$ \stackrel{03}{[-i]}\,\stackrel{12}{(+)}|
 \stackrel{56}{(+)}\,\stackrel{78}{[-]}
 ||\stackrel{9 \;10}{(+)}\;\;\stackrel{11\;12}{[-]}\;\;\stackrel{13\;14}{[-]}$
 &-1&$\frac{1}{2}$&-1&$\frac{1}{2}$&0&$-\frac{1}{2}$&$0$\\
 \hline
 8&$\nu_{L}$&$\stackrel{03}{(+i)}\,\stackrel{12}{[-]}|\stackrel{56}{(+)}\,\stackrel{78}{[-]}
 ||\stackrel{9 \;10}{(+)}\;\;\stackrel{11\;12}{[-]}\;\;\stackrel{13\;14}{[-]}$
 &-1&$-\frac{1}{2}$&-1&$\frac{1}{2}$&0&$-\frac{1}{2}$&$0$\\
 \hline\hline
 \end{tabular}
 \end{center}
 \caption{\label{Table Il.} The 8-plet of leptons - the members of $SO(1,7)$ subgroup of the 
 group $SO(1,13)$, 
 belonging to one Weyl left 
 handed ($\Gamma^{(1,13)} = -1 = \Gamma^{(1,7)} \times \Gamma^{(6)}$) spinor representation of 
 $SO(1,13)$. 
 It contains the colour chargeless left handed weak charged leptons and the right handed weak 
 chargeless leptons. The rest of notation is the same as in Table~\ref{Table Il.}.  
 }
 \end{table}

 In both tables the vectors are chosen to be the eigenvectors of the operators of 
 handedness $\Gamma^{(n)}$,  
 the generators $\tau^{13}, \, \tau^{23}, \,\tau^{33}$  $ \tau^{38}$,  $Y= \tau^{4} + \tau^{23}$ and
 $Q= Y + \tau^{13}$. They are also 
 eigenvectors of the corresponding $\tilde{S}^{ab}$, $\tilde{\tau}^{Ai}, A=1,2,3$ and $\tilde{Y}, \tilde{Q}$. 
One easily sees that the right handed vectors (with respect to $SO(1,3)$ )  are weak ($SU(2)_{I}$) 
chargeless and are doublets with respect to the second $SU(2)_{II}$, while the left handed are  weak 
charged and singlets with respect to $SU(2)_{II}$. 

The generators $\tilde{S}^{ab}$ transform each member of a family into the same member of other 
$2^{\frac{8}{2}-1}$ families. 
 The eight families of the first 
 member of the eightplet of quarks from Table~\ref{Table I.}, for example, that is of the right 
 handed $u$-quark 
 of the spin $\frac{1}{2}$,  are presented in the left column of Table~\ref{Table II.}. 
 The corresponding right handed neutrinos, belonging to eight different families, are presented 
 in the right column of the same table. The $u$-quark member of the eight families and the $\nu$ 
 members of the same eight families
 are generated by $\tilde{S}^{cd}$, $c,d \in \{0,1,2,3,5,6,7,8\}$ from any starting family.
 \begin{table}
 \begin{center}
 \begin{tabular}{|r||c||c||c||c||}
 \hline
 $I_R$ & $u_{R}^{c1}$&
 $ \stackrel{03}{[+i]}\,\stackrel{12}{(+)}|\stackrel{56}{(+)}\,\stackrel{78}{[+]}||
 \stackrel{9 \;10}{(+)}\:\; \stackrel{11\;12}{[-]}\;\;\stackrel{13\;14}{[-]}$ & 
 $\nu_{R}$&
 $ \stackrel{03}{[+i]}\,\stackrel{12}{(+)}|\stackrel{56}{(+)}\,\stackrel{78}{[+]}||
 \stackrel{9 \;10}{(+)}\;\;\stackrel{11\;12}{(+)}\;\;\stackrel{13\;14}{(+)}$ 
 \\
 \hline
  $II_R$ & $u_{R}^{c1}$&
  $ \stackrel{03}{[+i]}\,\stackrel{12}{(+)}|\stackrel{56}{[+]}\,\stackrel{78}{(+)}||
  \stackrel{9 \;10}{(+)}\;\;\stackrel{11\;12}{[-]}\;\;\stackrel{13\;14}{[-]}$ & 
  $\nu_{R}$&
  $ \stackrel{03}{(+i)}\,\stackrel{12}{[+]}|\stackrel{56}{(+)}\,\stackrel{78}{[+]}||
  \stackrel{9 \;10}{(+)}\;\;\stackrel{11\;12}{(+)}\;\;\stackrel{13\;14}{(+)}$ 
 \\
 \hline
 $III_R$ & $u_{R}^{c1}$&
 $ \stackrel{03}{(+i)}\,\stackrel{12}{[+]}|\stackrel{56}{(+)}\,\stackrel{78}{[+]}||
 \stackrel{9 \;10}{(+)}\;\;\stackrel{11\;12}{[-]}\;\;\stackrel{13\;14}{[-]}$ & 
 $\nu_{R}$&
 $ \stackrel{03}{(+i)}\,\stackrel{12}{[+]}|\stackrel{56}{[+]}\,\stackrel{78}{(+)}||
 \stackrel{9 \;10}{(+)}\;\;\stackrel{11\;12}{(+)}\;\;\stackrel{13\;14}{(+)}$ 
 \\
 \hline
 $IV_R$ & $u_{R}^{c1}$&
 $ \stackrel{03}{(+i)}\,\stackrel{12}{[+]}|\stackrel{56}{[+]}\,\stackrel{78}{(+)}||
 \stackrel{9 \;10}{(+)}\;\;\stackrel{11\;12}{[-]}\;\;\stackrel{13\;14}{[-]}$ & 
 $\nu_{R}$&
 $ \stackrel{03}{[+i]}\,\stackrel{12}{(+)}|\stackrel{56}{[+]}\,\stackrel{78}{(+)}|| 
 \stackrel{9 \;10}{(+)}\;\;\stackrel{11\;12}{(+)}\;\;\stackrel{13\;14}{(+)}$ 
 \\
 \hline\hline\hline
 $V_R$ & $u_{R}^{c1}$&
 $ \stackrel{03}{(+i)}\,\stackrel{12}{(+)}|\stackrel{56}{(+)}\,\stackrel{78}{(+)} ||
 \stackrel{9 \;10}{(+)}\;\;\stackrel{11\;12}{[-]}\;\;\stackrel{13\;14}{[-]}$ & 
 $\nu_{R}$&
 $ \stackrel{03}{(+i)}\,\stackrel{12}{(+)}|\stackrel{56}{(+)}\,\stackrel{78}{(+)} ||
 \stackrel{9 \;10}{(+)}\;\;\stackrel{11\;12}{(+)}\;\;\stackrel{13\;14}{(+)}$ 
 \\
 \hline
 $VI_R$ & $u_{R}^{c1}$&
 $ \stackrel{03}{(+i)}\,\stackrel{12}{(+)}|\stackrel{56}{[+]}\,\stackrel{78}{[+]}|| 
 \stackrel{9 \;10}{(+)}\;\;\stackrel{11\;12}{[-]}\;\;\stackrel{13\;14}{[-]}$ & 
 $\nu_{R}$&
 $ \stackrel{03}{(+i)}\,\stackrel{12}{(+)}|\stackrel{56}{[+]}\,\stackrel{78}{[+]}||
 \stackrel{9 \;10}{(+)}\;\;\stackrel{11\;12}{(+)}\;\;\stackrel{13\;14}{(+)}$ 
 \\
 \hline
 $VII_R$ & $u_{R}^{c1}$&
 $\stackrel{03}{[+i]}\,\stackrel{12}{[+]}|\stackrel{56}{(+)}\,\stackrel{78}{(+)}|| 
 \stackrel{9 \;10}{(+)}\;\;\stackrel{11\;12}{[-]}\;\;\stackrel{13\;14}{[-]}$ & 
 $\nu_{R}$&
 $\stackrel{03}{[+i]}\,\stackrel{12}{[+]}|\stackrel{56}{(+)}\,\stackrel{78}{(+)}||
 \stackrel{9 \;10}{(+)}\;\;\stackrel{11\;12}{(+)}\;\;\stackrel{13\;14}{(+)}$ 
 \\
 \hline
 $VIII_R$ & $u_{R}^{c1}$&
 $ \stackrel{03}{[+i]}\,\stackrel{12}{[+]}|\stackrel{56}{[+]}\,\stackrel{78}{[+]}|| 
 \stackrel{9 \;10}{(+)}\;\;\stackrel{11\;12}{[-]}\;\;\stackrel{13\;14}{[-]}$ & 
 $\nu_{R}$&
 $ \stackrel{03}{[+i]}\,\stackrel{12}{[+]}|\stackrel{56}{[+]}\,\stackrel{78}{[+]}|| 
 \stackrel{9 \;10}{(+)}\;\;\stackrel{11\;12}{(+)}\;\;\stackrel{13\;14}{(+)}$ 
 \\
 \hline 
 \end{tabular}
 \end{center}
 \caption{\label{Table II.} Eight families of the right handed $u_R$ quark with the spin $\frac{1}{2}$, 
  the colour charge $\tau^{33}=1/2$, $\tau^{38}=1/(2\sqrt{3})$ and of the colourless right handed  
  neutrino $\nu_R$ of the spin $\frac{1}{2}$ are presented in the left and in the right column, 
  respectively.
  $S^{ab}, a,b \in \{0,1,2,3,5,6,7,8\}$ transform $u_{R}^{c1}$ of the spin $\frac{1}{2}$ and the 
  chosen colour $c1$ to all the members of the same colour: to the right handed $u_{R}^{c1}$ 
  of the spin $-\frac{1}{2}$, 
  to the left $u_{L}^{c1}$ of both spins ($\pm \frac{1}{2}$), to the right handed $d_{R}^{c1}$ of both spins 
  ($\pm \frac{1}{2}$) and to the left handed $ d_{L}^{c1}$ of both spins ($\pm \frac{1}{2}$). They transform 
  equivalently the right handed   neutrino $\nu_R$ of the spin $\frac{1}{2}$ to the right handed 
  $\nu_R$ of the spin ($-\frac{1}{2}$), to  $\nu_L$ of both spins, to $e_R$ of both spins and to 
  $e_L$ of both spins. $\tilde{S}^{ab}, a,b \in \{0,1,2,3,5,6,7,8\}$ transform a chosen member of one family 
  into the same member of all the eight families.}
 \end{table}

Let us present also the quantum numbers of  the families from Table~\ref{Table II.}. 
In Table~\ref{Table IV.} 
the handedness of the families
$\tilde{\Gamma}^{(1+3)}(= -4i \tilde{S}^{03} \tilde{S}^{12})$, 
$\tilde{S}^{03}_{L}, \tilde{S}^{12}_L$, $\tilde{S}^{03}_{R}, \tilde{S}^{12}_R$ (the diagonal matrices of 
$SO(1,3)$ ), $\tilde{\tau}^{13}$ 
(of one of the two $SU(2)_{I}$), $\tilde{\tau}^{23}$ (of the second 
$SU(2)_{II}$) are presented. 

%
 \begin{table}
 \begin{center}
 \begin{tabular}{|r||r|r|r|r|r|r|r|r|r|r|r||}
 \hline
$i$ & $\tilde{\Gamma}^{(1+3)}$& $\tilde{S}^{03}_{L}$&$ \tilde{S}^{12}_L$& $\tilde{S}^{03}_{R}$& 
$\tilde{S}^{12}_R$& $\tilde{\tau}^{13}$ & $\tilde{\tau}^{23}$&$\tilde{\tau}^{4}$ & 
$\tilde{Y}'$&$\tilde{Y}$&$\tilde{Q}$ \\
\hline
\hline
$1$& $-1$ &$ - \frac{i}{2}$ &$ \frac{1}{2}$& $0$& $0$& $  \frac{1}{2}$& $0$&$-\frac{1}{2}$&$0$&$-\frac{1}{2}$&$0$\\ 
\hline
$2$& $-1$ &$ -\frac{i}{2}$ &$  \frac{1}{2}$& $0$& $0$& $-\frac{1}{2}$& $0$&$-\frac{1}{2}$&$0$&$-\frac{1}{2}$&$-1$\\
\hline
$3$& $-1$ &$ \frac{i}{2}$ &$ - \frac{1}{2}$& $0$& $0$& $ \frac{1}{2}$& $0$&$-\frac{1}{2}$&$0$&$-\frac{1}{2}$&$0$\\
\hline
$4$& $-1$ &$ \frac{i}{2}$ &$ - \frac{1}{2}$& $0$& $0$& $-\frac{1}{2}$& $0$&$-\frac{1}{2}$&$0$&$-\frac{1}{2}$&$-1$\\
\hline\hline
$5$& $1 $ & $0$ & $0$& $\frac{i}{2}$ &$ \frac{1}{2}$& $0 $& $ \frac{1}{2}$&$-\frac{1}{2}$&$\frac{1}{2}$&$0$&$0$\\ 
\hline
$6$& $1 $ & $0$ & $0$& $\frac{i}{2}$ &$  \frac{1}{2}$& $0 $& $-\frac{1}{2}$&$-\frac{1}{2}$&$-\frac{1}{2}$&$-1$&$-1$\\
\hline
$7$& $1 $ & $0$ & $0$& $-\frac{i}{2}$ &$- \frac{1}{2}$& $0 $& $ \frac{1}{2}$&$-\frac{1}{2}$&$\frac{1}{2}$&$0$&$0$\\
\hline
$8$& $1 $ &  $0$& $0$& $- \frac{i}{2}$ &$-\frac{1}{2}$& $0 $& $-\frac{1}{2}$&$-\frac{1}{2}$&$-\frac{1}{2}
$&$-1$&$-1$\\
\hline
\hline
 \end{tabular}
 \end{center}
 \caption{\label{Table IV.}  The quantum numbers of each member of the eight families presented in 
 Table~\ref{Table II.} are presented: The handedness of the families 
 $\tilde{\Gamma}^{(1+3)}= -4i \tilde{S}^{03} \tilde{S}^{12}$, the left and right handed $SO(1,3)$ 
 quantum numbers 
$\tilde{S}^{03}_{L}, \tilde{S}^{12}_L$, $\tilde{S}^{03}_{R}, \tilde{S}^{12}_R$ (of $SO(1,3)$ group in the 
$\tilde{S}^{mn}$ sector), $\tilde{\tau}^{13}$ 
 of  $SU(2)_{I}$, $\tilde{\tau}^{23}$ of the second 
$SU(2)_{II}$, $\tilde{\tau}^4$, 
$\tilde{Y}'= \tilde{\tau}^{23} -  
\tilde{\tau}^4 \, \tan\tilde{\theta}_2$, taking $\tilde{\theta}^2=0$, $\tilde{Y}
=\tilde{\tau}^{4} + \tilde{\tau}^{23}$, 
$\tilde{Q}= \tilde{\tau}^{4} + \tilde{S}^{56}$.  See also the ref.~\cite{normaproc2010talk}.
}
\end{table}

We see in Table~\ref{Table IV.} that four of the eight families are singlets with respect to 
one of the two $SU(2)$ ($SU(2)_{I}$) groups determined by $\tilde{S}^{ab}$ and doublets with respect to the 
second $SU(2)$ ($SU(2)_{II}$), while the remaining four families are doublets with respect to the first 
$SU(2)_{I}$ and singlets with respect to the second $SU(2)_{II}$. When the first break  
appears, to which besides the vielbeins also the spin connections  contribute, we expect that 
if only one of the two $SU(2)$ subgroups of $SO(1,7) \times U(1)$ breaking into
$SO(1,3) \times SU(2)\times U(1)$ contributes in the break~\cite{normaproc2010talk}, namely that of the charges 
$\tilde{\tau}^{2 i}$, together with $\tilde{N}^{i}_{-} $, there will be four families massless and mass 
protected after this break, namely those, 
which are singlets with respect to $\vec{\tilde{\tau}}^{2}$ and with respect to $\tilde{N}^{i}_{-} $
(Table~\ref{Table IV.}), while for the other four families  the vacuum expectation values of the 
scalars (particular combinations of vielbeins $f^{\sigma}{}_{s}$, and spin connections 
$\tilde{\omega}_{abs}, s \in \{5,8\}$) will take care of the mass matrices on the tree level and beyond  
the tree level. 

 \subsection{Discrete symmetries of the theory unifying spin and charges and explaining families}
\label{cpt} 

Let us define the discrete operators of the parity ($ P$) and of the charge conjugation 
($C$).
\begin{eqnarray}
\label{cp}
P &=& \gamma^0 \, \gamma^8 \, I_{x}\, , \nonumber\\
C &=& \Pi_{Im\, \gamma^a}\, \gamma^a \, K\,.
\end{eqnarray}
$K$ means complex conjugation, while in our choice of matrix representation of the $\gamma^a$ 
matrices $\Pi_{Im\, \gamma^a}\, \gamma^a \,= \gamma^2 \gamma^5 \gamma^7  \gamma^9 \gamma^{11} \gamma^{13}$.

One can easily check that $P$ transforms  the $ u_{R}^{c1}$ from the first row in Table~\ref{Table I.} 
into the $ u_{L}^{c1}$ of the seventh row in the same table. The $CP$ transforms the fermion 
states of table~\ref{Table I.} into the corresponding states of antifermions: $ u_{R}^{c1}$ 
from the first row in table~\ref{Table I.} with the spin $\frac{1}{2}$, weak chargeless
and of the colour charge
($(\frac{1}{2}, \frac{1}{2 \sqrt{3}})$) into a right handed antiquark $ \bar{u}_{R}^{\bar{c1}}$, 
weak charged and of the colour charge ($(-\frac{1}{2}, -\frac{1}{2 \sqrt{3}})$) as presented 
in table~\ref{Table anti}.
 \begin{table}
 \begin{center}
 \begin{tabular}{|r|c||c||c|c||c|c|c||r|r|}
 \hline
 i&$$&$|^a\psi_i>$&$\Gamma^{(1,3)}$&$ S^{12}$&$\Gamma^{(4)}$&
 $\tau^{13}$&$\tau^{23}$&$Y$&$Q$\\
 \hline\hline
 && ${\rm Octet},\;\Gamma^{(1,7)} =-1,\;\Gamma^{(6)} = 1,$&&&&&&& \\
 && ${\rm of \; antiquarks}$&&&&&&&\\
 \hline\hline
 1&$ \bar{u}_{R}^{\bar{c1}}$&$ \stackrel{03}{[-i]}\,\stackrel{12}{[-]}|
 \stackrel{56}{[-]}\,\stackrel{78}{(+)}
 ||\stackrel{9 \;10}{[-]}\;\;\stackrel{11\;12}{(+)}\;\;\stackrel{13\;14}{(+)} $
 &1&$-\frac{1}{2}$&-1&$-\frac{1}{2}$&$0$&$-\frac{1}{6}$&$-\frac{2}{3}$\\
 \hline 
 2&$\bar{u}_{R}^{\bar{c1}}$&$\stackrel{03}{(+i)}\,\stackrel{12}{(+)}|
 \stackrel{56}{[-]}\,\stackrel{78}{(+)}
 ||\stackrel{9 \;10}{[-]}\;\;\stackrel{11\;12}{(+)}\;\;\stackrel{13\;14}{(+)}$
 &1&$ \frac{1}{2}$&-1&$-\frac{1}{2}$&$0$&$-\frac{1}{6}$&$-\frac{2}{3}$\\
 \hline
 3&$\bar{d}_{R}^{\bar{c1}}$&$\stackrel{03}{[-i]}\,\stackrel{12}{[-]}|
 \stackrel{56}{(+)}\,\stackrel{78}{[-]}
 ||\stackrel{9 \;10}{[-]}\;\;\stackrel{11\;12}{(+)}\;\;\stackrel{13\;14}{(+)}$
 &1&$-\frac{1}{2}$&-1&$ \frac{1}{2}$&$0$&$-\frac{1}{6}$&$\frac{1}{3}$\\
 \hline 
 4&$ \bar{d}_{R}^{\bar{c1}} $&$\stackrel{03}{(+i)}\,\stackrel{12}{(+)}|
 \stackrel{56}{(+)}\,\stackrel{78}{[-]}
 ||\stackrel{9 \;10}{[-]}\;\;\stackrel{11\;12}{(+)}\;\;\stackrel{13\;14}{(+)} $
 &1&$\frac{1}{2}$&-1&$\frac{1}{2}$&$0$&$-\frac{1}{6}$&$\frac{1}{3}$\\
 \hline
 5&$\bar{d}_{L}^{\bar{c1}}$&$\stackrel{03}{(+i)}\,\stackrel{12}{[-]}|
 \stackrel{56}{(+)}\,\stackrel{78}{(+)}
 ||\stackrel{9 \;10}{[-]}\;\;\stackrel{11\;12}{(+)}\;\;\stackrel{13\;14}{(+)}$
 &-1&$-\frac{1}{2}$&1&$0$&$\frac{1}{2}$&$\frac{1}{3}$&$\frac{1}{3}$\\
 \hline
 6&$\bar{d}_{L}^{\bar{c1}} $&$\stackrel{03}{[-i])}\,\stackrel{12}{(+)}|
 \stackrel{56}{(+)}\,\stackrel{78}{(+)}
 ||\stackrel{9 \;10}{[-]}\;\;\stackrel{11\;12}{(+)}\;\;\stackrel{13\;14}{(+)} $
 &-1&$\frac{1}{2}$&1&$0$&$\frac{1}{2}$&$\frac{1}{3}$&$\frac{1}{3}$\\
 \hline
 7&$ \bar{u}_{L}^{\bar{c1}}$&$\stackrel{03}{(+i)}\,\stackrel{12}{[-]}|
 \stackrel{56}{[-]}\,\stackrel{78}{[-]}
 ||\stackrel{9 \;10}{[-]}\;\;\stackrel{11\;12}{(+)}\;\;\stackrel{13\;14}{(+)}$
 &-1&$-\frac{1}{2}$&1&$0$&$-\frac{1}{2}$&$-\frac{2}{3}$&$-\frac{2}{3}$\\
 \hline
 8&$\bar{u}_{L}^{\bar{c1}}$&$\stackrel{03}{[-i]}\,\stackrel{12}{(+)}|
 \stackrel{56}{[-]}\,\stackrel{78}{[-]}
 ||\stackrel{9 \;10}{[-]}\;\;\stackrel{11\;12}{(+)}\;\;\stackrel{13\;14}{(+)}$
 &-1&$ \frac{1}{2}$&1&$0$&$-\frac{1}{2}$&$-\frac{2}{3}$&$-\frac{2}{3}$\\
 \hline\hline
 \end{tabular}
 \end{center}
 \caption{\label{Table anti} The 8-plet of antiquarks  to the quarks 
 obtained from Table~\ref{Table I.} by the $CP$ ($= \gamma^2 \gamma^5 \gamma^7  
 \gamma^9 \gamma^{11} \gamma^{13}K \gamma^0 \gamma^8 \, I_{x}$) conjugation.} 
 \end{table}
 \section{The fermion-antifermion asymmetry within the theory unifying spin and charges 
 and explaining families}
 \label{matterasymmetry}

 As said in the abstract, I shall here  follow the ideas from
 the references~\cite{gross,rubakovshaposhnikov,dinekusenko,tapeiling}. The difference from the 
 studies there in here is, as explained, in the number of  families (there are two decoupled 
 groups of four families and consequently two stable families), in the number of 
 gauge fields contributing to the phase transitions and in the types of the gauge 
 fields contributing to phase transitions.
 
 Let us assume that the fermion-antifermion asymmetry is zero, when the expanding 
 universe cools down to the 
 temperature below the unification scale of the spin and the charges that is to the 
 temperature below, let say, $10^{16}$ TeV, when there are eight massless families, 
 manifesting  the symmetry $SO(1,3) \times SU(2) \times SU(2) \times U(1) \times SU(3)$, and 
 distinguishing among themselves in the quantum numbers defined by $\tilde{S}^{ab}$.

 Then we must investigate, how much do the following processes contribute to  the fermion-antifermion 
 asymmetry  in non equilibrium thermal processes in the expanding universe:   
 
 \begin{itemize}
  
 \item The nonconservation of currents on the quantum level
 due to the triangle anomalies~\cite{tapeiling,rubakovshaposhnikov,dinekusenko}, 
 which are responsible for 
 $P$ and $CP$ nonconservation  
 \begin{eqnarray}
 \label{nonconsrvedcurrents}
 \partial^{m} \, j^{A i \,\alpha  (i)}_{ m} &=& \frac{(g^{A})^2}{8 \pi^2}  \; \frac{1}{2} \,
 \varepsilon_{mnpr}\, F^{Ai\, mn} F^{Ai \, pr}.
 \end{eqnarray}
 Here $j^{A i \,\alpha }_{ m} $ stays  for the currents of fermions (and antifermions), which carry 
 a particular charge denoted by a charge group $A$, in our case $A=4$ means the $U(1)$ 
 charge originating in $SO(6)$,   $A=3$ means the $SU(3)$ (colour) 
 charge, 
 $A=2_I$ means the weak $SU(2)_I$ charge of the left handed doublets, while $A=2_{II}$  stays 
  for the $SU(2)_{II}$
 charge of the right handed singlets before the $SU(2)_{II}$ break, 
 $A=1$ stays for the actual $U(1)$ charge (the {\it standard model} like hyper charge 
 after the $SU(2)_{II}$ break and the electromagnetic one after the weak break).

 In my case also the fields, which look like scalar fields in $d=(1+3)$, $\tilde{A}^{\tilde{A}i}_{s}$, 
 $s,t \in{5,6,\cdots}$, and to which the fermions are 
 coupled, contribute. 
 
 All the fermions and antifermions, which are coupled to a particular gauge 
 field $A^{Ai}_{m}$ and 
 in my case  also $\tilde{A}^{\tilde{A}i}_{s}$ contribute to the current 
 \begin{eqnarray}
 \label{current}
 j^{A i \,\alpha (i)}_m = \psi^{Ai\, \alpha (i)\dagger}\, \gamma^0 \gamma^m \,\psi^{Ai \, \alpha (i)}. 
 \end{eqnarray}
 $(i) \in  \{1,8\}$ enumerates families, in my case twice four families which are distinguishable 
 by the quantum numbers originating in $\tilde{S}^{ab}$, namely, after the break of $SU(2)_I$  
 the lower four families, which are 
 doublets with respect to $\tilde{N}^{i}_{+}$ and $\tilde{\tau}^{I\,i}$ and singlets with respect to 
 $\tilde{N}^{i}_{-}$ and $\tilde{\tau}^{II\,i}$,  stay massless, 
 while the upper four families are doublets 
 with respect to  $\tilde{N}^{i}_{-}$ and $\tilde{\tau}^{IIi}$ and singlets with respect to 
 $\tilde{N}^{i}_{+}$ and $\tilde{\tau}^{Ii}$. After the electroweak break all the eight families 
 become massive, but the upper four families have no mixing matrix elements since the way of breaking 
 leaves all the 
 $\omega_{msa}$ and $\tilde{\omega}_{msa}$, with $m=0,1,2,3; s=5,6,\cdots$, equal to zero. 
  $\alpha$  distinguishes the multiplets in each family, in my case of the two $SU(2)$ 
  gauge groups $\alpha$ distinguishes the $SU(2)_I$ doublets, that is one colour singlet and 
  one colour triplet, 
  and the $SU(2)_{II}$ doublets, again one colour singlet and one colour triplet.
  $A^{Ai}_{m}$ are the corresponding gauge fields, with tensors  
 $F^{A}_{mn} =  \tau^{A i} \,F^{A i}_{mn}$ and $F^{A i}_{mn}= A^{A i}_{n,m} - A^{A i}_{m,n} +
 g^A \, f^{Ai j k} \, A^{A j}_{m}\, A^{A k}_{n} $. (The scalar fields 
 $\tilde{A}^{\tilde{A}i}_{s}$ define tensors  
  $\tilde{F}^{\tilde{A}i\, st}= \tilde{A}^{\tilde{A}i}_{t,s} -
  \tilde{A}^{\tilde{A}i}_{s,t} + g^{\tilde{}A} \, f^{A\, ijk} \tilde{A}^{\tilde{A}j}_{s}\,
 \tilde{A}^{\tilde{A}k}_{t}$.)   
 
 The nonconserved currents affect the fermions and antifermions. (In the later case the 
 $\tau^{A i}$ are replaced by $\bar{\tau}^{A i}$, both fulfilling the same 
 commutation relations $\{\tau^{A i}, \tau^{B j} \}_{-} = 
 i \delta^{AB} \, f^{A \,ijk} \, \tau^{A k} $, $\{\bar{\tau}^{A i}, \bar{\tau}^{B j} \}_{-} = 
 i \delta^{AB} \, f^{A \,ijk} \, \bar{\tau}^{A k}$). One obtains $\bar{\tau}^{A i}$ from 
$\tau^{A i}$ by the 
 $CP$ transformation $P= \gamma^0 \gamma^8 \,I_{x}$, while $C= \prod_{ Im \gamma^a}\,\gamma^a \, K$
 (See~\ref{cpt}).

 \item The nonconservation of the fermion numbers originating in the 
 complex phases of  the mixing matrices of the two times $4 \times 4$ mass matrices for 
 each member of a family, after the two successive breaks causes two phase transitions 
 when the symmetry $SU(2) \times SU(2) \times U(1)$ breaks first to $SU(2) \times U(1)$ 
 and finally to  $U(1)$ and the two types of gauge 
 fields manifest their masses while the two groups of four with the 
 mixing matrices decoupled families gain nonzero mass matrices in the first break the 
 upper four families and in the second break the lower four families.  
 
 \end{itemize}

I am following here the references~\cite{gross,rubakovshaposhnikov,dinekusenko,tapeiling}.
The nonconservation of currents may be expected whenever the  non-Abelian  gauge fields 
 manifest a non trivial structure of  vacua, originating 
in  the {\it instanton} solutions of the Euclidean non-Abelian gauge theories in $(1+3)-$dimensional 
space, that is in $A^{A}_{m}$, which  fulfil the boundary condition 
$\lim_{r \to \infty} \;\tau^{Ai}\,A^{Ai}_{m} = U^{-1} \partial_m U$, summed over $i$ for a particular 
gauge group $A$ (and 
similarly might be that the fields 
$ \lim_{\rho \to \infty} \;\tilde{\tau}^{\tilde{A}i}\,\tilde{A}^{\tilde{A}i}_{s}= 
U^{-1} \partial_s U$, with $r=\sqrt{(x^0)^2 + \vec{x}^2}$ and $\rho = \sqrt{\sum_s}\, (x^s)^2$, 
for a particular $\tilde{A}$ , contribute as well, where the effect of the triangle anomalies 
in the case of scalar gauge 
fields depending on $x^{\sigma}, \sigma = 5,6,7,8$ and the corresponding meaning of the winding numbers  
distinguishing among the different vacua in this case might be negligible and should be studied). 
The vacua  distinguish among themselves in the topological quantum numbers $n_{A}$ ($ n_{\tilde{A}}$), 
determined by  a particular choice of $U$
\begin{eqnarray}
\label{na}
n_{A} = \frac{(g^{A})^2 }{16 \pi^2} \, \int d^4 x \, \varepsilon_{mnpr}\, Tr (F^{Amn} F^{Apr})= 
\frac{(g^{A})^2}{32 \pi^2} \, \int d^{4} x \partial_{m} K^{A\, m}, 
\end{eqnarray}
where $K^{A}_m= \sum_{i}4 \varepsilon_{mnpr} \,( A^{Ai}_{n} \partial_{p} A^{A}_{r}+ 
\frac{2}{3}\, g^A f^{Aijk} A^{Ai}_{n} A^{Aj}_{p}A^{Ak}_{r}).$ 
(Similarly also the topological quantum number $n_{\tilde{A}}$ might be non negligible.)

Instanton solutions fulfilling 
the boundary condition $\lim_{r \to \infty}\; \tau^{Ai}\, A^{Ai}_{m}  = U^{-1} \partial_m U$ 
for a particular gauge group $A$ 
(or $\lim_{\rho \to \infty}\; \tilde{\tau}^{\tilde{A}i}\,\tilde{A}^{\tilde{A}i}_{s} = 
U^{-1} \partial_m U$ for a particular $\tilde{A}$), each with its own $U$ for a particular 
$A$ (or $\tilde{A}$), connect vacua  
$|n_{A}>$ with different winding numbers~\footnote{In the ref.~\cite{tapeiling,gross} the 
{\it instanton} field 
$\tau^{Ai}\, A^{Ai}_{m}= \frac{r^2}{r^2 + \lambda^2} \,U^{-1} \partial_m U$, with $U = 
\frac{x^0 + i\vec{\sigma}^{A} \cdot \vec{x}}{r}$, $r^2 = (x^0)^2 + \vec{x}^2,$ is presented. 
Operators $\vec{\sigma}^{A}$ 
are the three Pauli matrices,  used  to denote the $SU(2)$ gauge group in this case: $\vec{\tau}^{A}=
\frac{\vec{\sigma}^{A}}{2}$, $\{\tau^{Ai}, \tau^{Aj}\}_{-} = i \varepsilon^{ijk} \tau^{Ak}$. 
The corresponding 
action  $ \int d^{4}x\, \frac{1}{2} \varepsilon_{mnpr} \, F^{Ai\, mn} F^{Ai pr}= \frac{8 \pi^2}{g^2}$, 
while  $U = 
\frac{x^0 + i\vec{\sigma}^{A} \cdot \vec{x}}{r}$, defines the $n=1$ vacuum state. 
}
$n_{A}$ (and correspondingly for $n_{\tilde{A}}$).
The true vacuum $|\theta^{A}>$ is for each $A$ (let it count also $\tilde{A}$) in a stationary 
situation a superposition of the vacua, 
determined by the time 
independent gauge transformation~\cite{gross} ${\cal T}$, 
${\cal T} |\theta^{A}> = e^{i\theta^{A}} |\theta^{A}>$, 
where $\theta^{A}$ is a parameter, which weights  
 the contribution of a vacuum to the effective Lagrange density 
${\cal L}_{eff}= {\cal L} + \sum_{A} \, \frac{\theta^{A}}{16 \pi^2} \, 
F^{Ai \,mn} \frac{1}{2} \varepsilon_{mnpr}
F^{Ai\, pr}$, for a particular gauge field. ${\cal T}$ acts as the raising operator for the handedness 
(chirality).  
The second term of the effective Lagrange density ${\cal L}_{eff}$  violates parity $P$ and  then also $CP$.  
The vacuum state with the definite handedness has also a definite topological quantum number. 
In the presence of the massless fermions all the vacua  $|\theta^{A}>$, for each $A$,  are equivalent.

The fermion currents~(Eq.(\ref{current})) are not conserved in  processes, for which 
 the gauge fields are such  that the corresponding winding number $n_A$ of Eq.~(\ref{na}) is nonzero.
 Correspondingly also the fermion (and antifermion numbers), carrying the corresponding charge, are 
 not conserved 
\begin{eqnarray}
\label{nafermion}
\Delta n_{A i \,\alpha  (i)}= n_{A}. 
\end{eqnarray}
The fermion number of all the fermions interacting with the same  non-Abelian gauge field with nonzero 
winding number, either of a vector or of a scalar type (whose contribution should be studied and 
hopefully understood),  changes 
in such processes for the {\it same amount}: Any member of a family, interacting with the particular 
field and therefore also the corresponding members of each family, either a quark or  a lepton member 
of doublets, change for the same amount, before the  breaks or after the breaks 
(in my case first from $SO(1,3) \times SU(2) \times SU(2) \times U(1) \times SU(3)$ to
$SO(1,3) \times SU(2)  \times U(1) \times SU(3)$ and finally to $SO(1,3)  \times U(1) \times SU(3)$) 
of the symmetries.

For a baryon three quarks are needed. It is the conservation of the colour charge which 
requires  that  the lepton number and the baryon number ought to be conserved 
separately as long as the charge group is a global symmetry. The transformations, which allow 
rotations of a lepton to a quark or opposite, conserve the fermion number, but not the lepton and not the 
baryon number.

  Instanton solutions of the non-Abelian gauge fields, which connect different vacua (see the 
refs.~\cite{tapeiling}, page 481, and~\cite{rubakovshaposhnikov}, page 6), are characterized by the 
 highest value of the {\it instanton} field between the two vacua, that is by the {\it sphaleron} energy. 

The question arises, can the {\it instanton} solutions be responsible for the baryon asymmetry of the universe?
The authors of the papers~\cite{rubakovshaposhnikov,dinekusenko} discuss and evaluate the 
probability for tunnelling from one vacuum to the other at low energy regime and also at the energies of 
{\it sphalerons}. 
When once the  system of gauge fields is in one vacuum the probability for the 
transition to another vacuum depends not only on the {\it sphalerons} height (energy)  
but also on the temperature. If the temperature is low, then the 
transition is negligible. At the temperature above the phase transition 
(the authors~\cite{rubakovshaposhnikov}
discuss the electroweak phase transition starting at around $100$ GeV, while in my case there is also 
the $SU(2)_{II}$ phase transition at around $10^{16}$ GeV or slightly below) when the fermions 
are massless and the
expansion rate of the universe is much slower that the rate of nonconservation of the fermion number,
and in the case of non equilibrium processes in phase transitions, the fermion number nonconservation 
can be large. 
The authors conclude that more precise  evaluations (treating several models) of the probability 
that in a non thermal equilibrium   phase  transition and below it the fermion  
number would not be conserved   
due to  transitions to vacua with different winding numbers 
in the amount as observed  for the (first family) baryon number excess in the universe are needed.

What can be concluded about the fermion number asymmetry, caused by the transitions of gauge fields 
to different vacua, in my case, where at energies above the 
$ SU{2}_{II}$ phase transition there are eight families of massless fermions, with the charges manifesting 
the symmetries first of $SU(2)_I \times SU(2)_{II} \times U(1)$ and correspondingly with the two 
kinds of the vector gauge 
$SU(2)$ fields which both might demonstrate  the vacua with different winding numbers? In addition 
also the scalar gauge fields might contribute with their even more rich vacua (if they do that at all). 
The phase transitions caused first by the break of the symmetry $SU(2)_I \times SU(2)_{II} \times U(1)$ to 
$SU(2)_{II} \times U(1)$, when the upper four families gain masses (and the  corresponding gauge 
vector fields become massive)
and then by the final break to $U(1)$, with the $\tilde{S}^{ab}$ sector causing the masses 
in both transitions and may be 
also taking care of the richness of vacua with different winding numbers, 
might show up after a careful study as a mechanism 
for generating the fermion-antifermion (or the antifermion-fermion) asymmetry. Although I do not 
yet see, 
how do the non equilibrium processes in the first order phase transitions decide about the 
excess of either fermions or  of antifermions.

So, is it in my case possible that the two successive non equilibrium phase transitions 
leave the excess of antifermions in the case of the upper four families and the excess  
of fermions in the lower four families? Or there is a negligible excess of either fermions or 
antifermions in the upper four families? We saw in the ref.~\cite{gn} that an excess of either 
fermions or antifermions is not important for massive enough (few $100$ TeV)  stable 
fifth family members. The excess of fermions over antifermions is certainly what universe made 
a choice of for the lower four families, whatever the reason for this fact is. Can this be easier 
understood  within the {\it spin-charge-family-theory}?
All these need a careful study.

The fermion number nonconservation  originates also in the complex phases of the mass mixing matrices 
of  each of the two groups of four family members. It  might be that the vacua, triggered by  
{\it instanton} solutions of the gauge vector and scalar fields, and the mass matrices, determined 
on the tree level by the 
vacuum expectation values of the scalar gauge fields in the $\tilde{S}^{ab}$ sector, are connected  
(since in the {\it instanton} case also the 
scalar fields, the gauge fields of charges originating in $\tilde{S}^{ab}$ might exhibit the 
{\it instanton} solutions).

\section{Conclusion}
\label{conclusion}

In this contribution I pay attention to the origin of baryon asymmetry of our universe within the 
{\it spin-charge-family-theory} under the assumption that  the asymmetry is caused {\bf i.} by the 
{\it instanton} solutions of the non-Abelian gauge fields which determine vacua with different 
winding numbers and {\bf ii.} by the complex matrix elements of the mixing matrices. 

The {\it spin-charge-family-theory} namely assumes besides the Dirac Clifford algebra objects also the 
second ones $\tilde{\gamma^a}$ as a necessary mechanism (or better a mathematical tool) 
which should be used in order that we consistently describe both: spin and charges, as well as families.
The second kind is namely responsible for generating families, defining the equivalent 
representations with respect to the Dirac one. Correspondingly there are 
besides  the two kinds of the vector gauge fields, the  $SU(2)_{I}$ and $SU(2)_{II}$, also the scalar 
gauge fields, the two $SU(2)$ from $SO(4)$ and the two $SU(2)$ from $SO(1,3)$, the superposition of the 
gauge fields of 
$\tilde{S}^{ab} (= \frac{i}{4} (\tilde{\gamma^a} \tilde{\gamma^b} - \tilde{\gamma^b}\tilde{\gamma^a}))$,
 which might contribute to vacua  with different winding numbers (what has to be studied). The scalar fields,  
originating in the $\tilde{S}^{ab}$ charges, are responsible  with their  vacuum expectation values 
(and in loop corrections)
for the mass matrices of fermions after the breaks of symmetries.

The theory predicts twice four families (which 
differ in the family quantum numbers in the way that the upper four families are doublets 
with respect to $\tilde{\tau}^{II \, i}$ and $\tilde{N}^{i}_{-}$, while the lower four families are 
doublets with respect to $\tilde{\tau}^{I \, i},$ and $\tilde{N}^{i}_{+}$) which all  are 
massless above the last two phase transitions. 

What should be clarified in the {\it spin-charge-family-theory} is whether the predicted twice four  
families (rather than once three families of the {\it standard model}) and the fact that there are gauge 
fields belonging to two kinds of generator ($S^{ab}$ and $\tilde{S}^{ab}$) make the baryon number 
asymmetry easier to be understood within these two phenomena --- the {\it instanton} responsibility for the
fermion number nonconservation and the complex matrix elements of the mixing matrices responsibility 
for the fermion number nonconservation.

The manifestation of the {\it instanton} gauge vector and scalar fields in the determination of the 
properties of the vacuum might be correlated with 
the vacuum expectation values of the scalar fields defining the mass matrices of  twice the four families.
 Both manifestations appear in possibly non equilibrium phase transitions of the expanding universe, 
which cause breaking of particular symmetries and also the 
fermion number nonconservation. In this contribution I just follow the way 
suggested by the ref.~\cite{rubakovshaposhnikov} 
and by the authors cited in this reference, while taking into account the requirement of the 
{\it spin-charge-family-theory}. The fermion number nonconservation obviously ended in the   
excess of (what we call) fermions for the lower four families, while for the upper four 
families we have to see whether there is the  
excess of either the stable fifth family  fermions or antifermions.  
To answer these questions a careful study is needed. It even might be that there was at the non equilibrium 
phase transitions the same  excess of antifermions for the upper four families as it is of fermions 
for the  lower four families, while later the complex matrix elements in the mixing matrices change 
this equality drastically. But yet it must be understood the origin of both sources of the 
fermion number nonconservation.

\appendix*

\section{Some useful relations}
\label{sabprop}

The following Cartan subalgebra set of the algebra $S^{ab}$ (for both sectors) is chosen:
\begin{eqnarray}
S^{03}, S^{12}, S^{56}, S^{78}, S^{9 \;10}, S^{11\;12}, S^{13\; 14}\nonumber\\
\tilde{S}^{03}, \tilde{S}^{12}, \tilde{S}^{56}, \tilde{S}^{78}, \tilde{S}^{9 \;10}, 
\tilde{S}^{11\;12}, \tilde{S}^{13\; 14}.
\label{cartan}
\end{eqnarray}
%
A left handed ($\Gamma^{(1,13)} =-1$) eigen state of all the members of the 
Cartan  subalgebra 
\begin{eqnarray}
&& \stackrel{03}{(+i)}\stackrel{12}{(+)}|\stackrel{56}{(+)}\stackrel{78}{(+)}
||\stackrel{9 \;10}{(+)}\stackrel{11\;12}{(-)}\stackrel{13\;14}{(-)} |\psi \rangle = \nonumber\\
&&\frac{1}{2^7} 
(\gamma^0 -\gamma^3)(\gamma^1 +i \gamma^2)| (\gamma^5 + i\gamma^6)(\gamma^7 +i \gamma^8)||
\nonumber\\
&& (\gamma^9 +i\gamma^{10})(\gamma^{11} -i \gamma^{12})(\gamma^{13}-i\gamma^{14})
|\psi \rangle .
\label{start}
\end{eqnarray}
represent the $u_R$-quark with spin up and of one colour.

$ \tilde{S}^{ab} $  generate families from the starting $u_R$ quark
In particular $\tilde{S}^{03}(= \frac{i}{2}
[\stackrel{03}{\tilde{(+i)}} \stackrel{12}{\tilde{(+)}} +
\stackrel{03}{\tilde{(-i)}} \stackrel{12}{\tilde{(+)}} +
\stackrel{03}{\tilde{(+i)}} \stackrel{12}{\tilde{(-)}}+
\stackrel{03}{\tilde{(-i)}} \stackrel{12}{\tilde{(-)}}])$  applied on 
a right handed $u_R$-quark with spin up and a particular colour generate a state which is again 
 a right handed $u$-quark of the same colour.
\begin{eqnarray}
\stackrel{03}{\tilde{(-i)}}\stackrel{12}{\tilde{(-)}} &&
\stackrel{03}{(+i)}\stackrel{12}{(+)}| \stackrel{56}{(+)} \stackrel{78}{(+)}||
\stackrel{9 10}{(+)} \stackrel{11 12}{(-)} \stackrel{13 14}{(-)}=\nonumber\\
&&\stackrel{03}{[\,+i]} \stackrel{12}{[\,+\,]}| \stackrel{56}{(+)} \stackrel{78}{(+)}||
\stackrel{9 10}{(+)} \stackrel{11 12}{(-)} \stackrel{13 14}{(-)},
\end{eqnarray}
where 
\begin{eqnarray}
\stackrel{ab}{(\pm i)}         &=& 
\frac{1}{2}\, ( \gamma^a \mp  \gamma^b), 
\stackrel{ab}{(\pm 1)} =          \frac{1}{2} \,( \gamma^a \pm i\gamma^b),\nonumber\\
\stackrel{ab}{[\pm i]}& =& \frac{1}{2} (1 \pm   \gamma^a \gamma^b), \quad
\stackrel{ab}{[\pm 1]} = \frac{1}{2} (1 \pm i \gamma^a \gamma^b), \nonumber\\
\stackrel{ab}{\tilde{(\pm i)}} &=& 
\frac{1}{2}  (\tilde{\gamma}^a \mp  \tilde{\gamma}^b), \quad
\stackrel{ab}{\tilde{(\pm 1)}} = 
\frac{1}{2}  (\tilde{\gamma}^a \pm i\tilde{\gamma}^b), \nonumber\\ 
\stackrel{ab}{\tilde{[\pm i]}} &=& \frac{1}{2} (1 \pm \tilde{\gamma}^a \tilde{\gamma}^b), \quad
\stackrel{ab}{\tilde{[\pm 1]}} = \frac{1}{2} (1 \pm i \tilde{\gamma}^a \tilde{\gamma}^b). 
\label{deftildefun}
\end{eqnarray}

We present below some useful relations which are easy to derive~\cite{pikanorma}. 

\begin{eqnarray}
\label{relations}
\stackrel{ab}{(k)} \stackrel{ab}{(k)}& =& 0, \;   \stackrel{ab}{(k)} \stackrel{ab}{(-k)}
= \eta^{aa}  \stackrel{ab}{[k]}, \; 
\stackrel{ab}{[k]} \stackrel{ab}{[ k]} =   \stackrel{ab}{[k]}, \nonumber\\
\stackrel{ab}{[k]} \stackrel{ab}{[-k]} &=& 0, 
\stackrel{ab}{(k)} \stackrel{ab}{[ k]} = 0,\quad \; 
\stackrel{ab}{[k]} \stackrel{ab}{( k)}  = \stackrel{ab}{(k)}, \nonumber\\ 
\stackrel{ab}{(k)} \stackrel{ab}{[-k]} &=& \stackrel{ab}{(k)}\, ,
\quad \; \stackrel{ab}{[k]} \stackrel{ab}{(-k)} =0.   
\end{eqnarray}
\begin{eqnarray}
\stackrel{ab}{\tilde{(k)} }  \stackrel{ab}{(k)}& =& 0, 
\quad \;
 \stackrel{ab}{\tilde{(-k)}} \stackrel{ab}{(k)}= 
-i \eta^{aa}                 \stackrel{ab}{[k]},\nonumber\\ 
 \stackrel{ab}{\tilde{( k)}} \stackrel{ab}{[k]}&=& 
i  \stackrel{ab}{(k)},\quad
 \stackrel{ab}{\tilde{( k)}} \stackrel{ab}{[-k]} = 0.
\label{graphbinomsfamilies}
\end{eqnarray}
\begin{eqnarray}
N^{\pm}_{+}         &=& N^{1}_{+} \pm i \,N^{2}_{+} = 
 - \stackrel{03}{(\mp i)} \stackrel{12}{(\pm )}\,, \quad N^{\pm}_{-}= N^{1}_{-} \pm i\,N^{2}_{-} = 
  \stackrel{03}{(\pm i)} \stackrel{12}{(\pm )}\,,\nonumber\\
\tilde{N}^{\pm}_{+} &=& - \stackrel{03}{\tilde{(\mp i)}} \stackrel{12}{\tilde{(\pm )}}\,, \quad 
\tilde{N}^{\pm}_{-}= 
  \stackrel{03} {\tilde{(\pm i)}} \stackrel{12} {\tilde{(\pm )}}\,,\nonumber\\ 
\tau^{1\pm}         &=& (\mp)\, \stackrel{56}{(\pm )} \stackrel{78}{(\mp )} \,, \quad   
\tau^{2\pm}=            (\mp)\, \stackrel{56}{(\mp )} \stackrel{78}{(\mp )} \,,\nonumber\\ 
\tilde{\tau}^{1\pm} &=& (\mp)\, \stackrel{56}{\tilde{(\pm )}} \stackrel{78}{\tilde{(\mp )}}\,,\quad   
\tilde{\tau}^{2\pm}= (\mp)\, \stackrel{56}{\tilde{(\mp )}} \stackrel{78}{\tilde{(\mp )}}\,.
\end{eqnarray}
 \end{document}